\documentclass[twocolumn,prl,amsmath,amssymb,floatfix,superscriptaddress]{revtex4-2} 
\usepackage{amsmath}
\usepackage{amssymb}
\usepackage{graphicx}
\usepackage{bm}
\usepackage{color}
\usepackage{hyperref}
\usepackage{graphicx}
\usepackage{upgreek}
\usepackage{scalerel}
\usepackage{multirow}
 
\usepackage{pgffor}
\usepackage{pdfpages}
\usepackage{mathdots}
\usepackage{float}
\usepackage{silence}
\usepackage{physics} 
\usepackage{lscape}   
\usepackage{color, colortbl}
\usepackage[table]{xcolor}
\usepackage{comment}  
\makeatletter
\AtBeginDocument{\let\LS@rot\@undefined}
\makeatother

\begin{document}

\title{Enhancing the Josephson diode effect with Majorana bound states}
\author{Jorge Cayao}
\affiliation{Department of Physics and Astronomy, Uppsala University, Box 516, S-751 20 Uppsala, Sweden}
\author{Naoto Nagaosa}
\affiliation{Center for Emergent Matter Science (CEMS), RIKEN, Wako, Saitama, 351-0198, Japan}

\author{Yukio Tanaka}
\affiliation{Department of Applied Physics, Nagoya University, Nagoya 464--8603, Japan}
\affiliation{Research Center for Crystalline Materials Engineering, Nagoya University, Nagoya 464-8603, Japan}

\date{\today}
\begin{abstract}
We consider phase-biased  Josephson junctions with spin-orbit coupling under external magnetic fields and study the emergence of the Josephson diode effect in the presence of Majorana bound states. We show that  junctions having middle regions with Zeeman fields along the spin-orbit axis  develop  a low-energy Andreev spectrum that is asymmetric with respect to the superconducting phase difference $\phi=\pi$, which is strongly influenced by Majorana bound states in the topological phase. This asymmetric Andreev spectrum gives rise to anomalous current-phase curves and critical currents  that are different for positive and negative supercurrents, thus signaling the emergence of the Josephson diode effect. While this effect exists even in the trivial phase, it gets   enhanced in the topological phase  due to the spatial nonlocality   of Majorana bound states. Our work thus establishes the utilization of topological superconductivity for enhancing the functionalities of Josephson diodes.
\end{abstract}
\maketitle

Diodes are devices that conduct current primarily along one direction and constitute key building blocks of numerous electronic components \cite{coldren2012diode,mehdi2017thz,semple2017flexible,tokura2018nonreciprocal,kim2020analogue,loganathan2022rapid}.  Diodes have been initially studied in the normal state \cite{braun1875ueber,sze2008semiconductor}, where inevitable energy losses appear due to finite resistance. This issue has been resolved in superconductors (Ss), which showed superior diode functionalities \cite{wakatsuki2017nonreciprocal,hoshino2018nonreciprocal,yasuda2019nonreciprocal}, with  dissipationless supercurrents in   bulk systems \cite{ando2020observation,PhysRevLett.128.037001,PhysRevLett.131.027001,PhysRevB.106.104501,he2022phenomenological,yuan2022supercurrent,lin2022zero,scammell2022theory,banerjee2023enhanced} and in  Josephson junctions (JJs) \cite{wu2022field,baumgartner2022supercurrent,pal2022josephson,davydova2022universal,PhysRevLett.129.267702,PhysRevB.108.054522,PhysRevLett.130.266003,PhysRevB.107.245415}. Of particular interest are  diodes in JJs,  known as Josephson diodes (JDs),  because supercurrents here are   controlled by virtue of the Josephson effect, which arises due to the finite phase difference between coupled Ss \cite{RevModPhys.51.101,Tinkham}.  This  phase also enables the formation of Andreev bound states (ABSs), which   determine  the profile of  supercurrents  \cite{kulik1975,furusaki1991dc,PhysRevB.45.10563,Beenakker:92,Furusaki_1999,kashiwaya2000tunnelling,PhysRevB.64.224515,PhysRevLett.96.097007,RevModPhys.76.411,sauls2018andreev,mizushima2018multifaceted} and   reveal the nature of  emergent superconductivity \cite{tanaka2011symmetry,tiira17,PhysRevLett.121.047001,PhysRevX.9.011010,ren2019topological,Fornieri_2019,PhysRevLett.124.226801,PhysRevLett.125.116803,lutchyn2018majorana,prada2019andreev,frolov2019quest,flensberg2021engineered}.
 
 \begin{figure}[!t]
\centering
	\includegraphics[width=0.99\columnwidth]{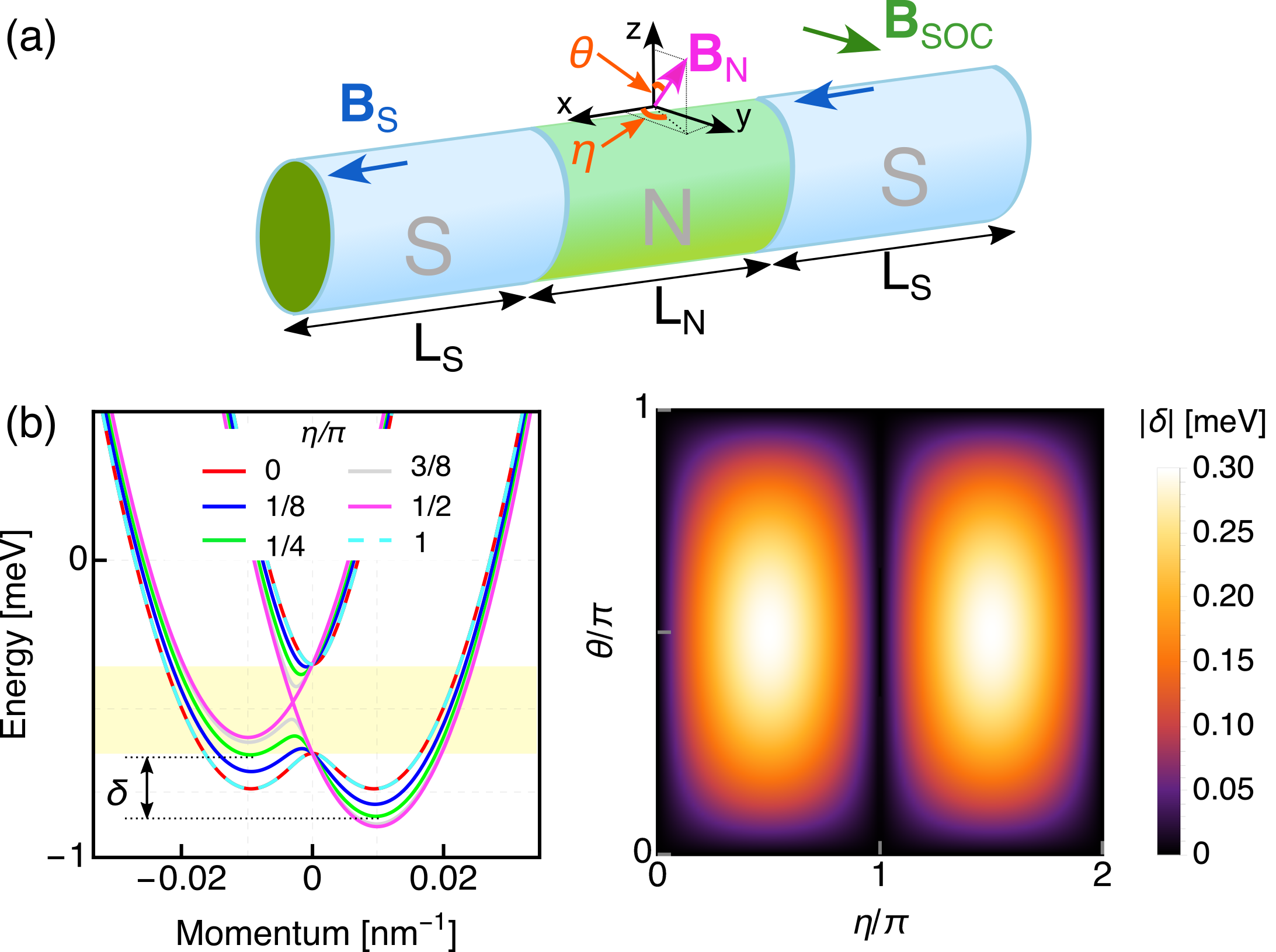}\\
 	\caption{(a) A  JJ based on a  nanowire with SOC field ${\mathbf B}_{\rm SOC}$ along $y$-axis (green), with the  N(S) region of length $L_{\rm N(S)}$.  The N (S)  region has a Zeeman field $\mathbf{B_{\rm N(S)}}$  with components perpendicular and parallel (only perpendicular) to ${\mathbf B}_{\rm SOC}$, see magenta (blue) arrow. (b) Left panel: Energy versus momentum without superconductivity: while a perpendicular Zeeman field  opens a gap at zero momentum (yellow region), a parallel  term  induces an asymmetry in the  bands, $\delta$, seen by fixing the angle  of $\mathbf{B_{\rm N}}$ with the $z$-axis to $\theta=\pi/2$ and varying the   angle from the $x$-axis $\eta$; $\delta$ is  shown for $\eta = \pi/4$.  The solid red ($\eta=0$) and  dashed cyan curves ($\eta=\pi$) are superimposed.  Right panel: $\delta$ at the SOC momenta as a function of $\theta$ and $\eta$. Parameters: $B_{\rm N}=0.15$meV, $\alpha=40$meVnm, $\mu=0.5$meV.}
\label{Fig0} 
\end{figure}

Most of the studies on JDs have involved systems with spin-orbit coupling (SOC) and magnetic fields, with  experiments that strongly support their realization in    semiconductor-S junctions \cite{baumgartner2022supercurrent,mazur2022gatetunable,PhysRevLett.131.027001,nadeem2023superconducting,valentini2023parityconserving}.  
Moreover, JDs are characterized based on their quality factors, which measure  the diode's ability  to conduct current along one direction. In this regard, recent studies have reported sizable and highly tunable quality factors, showing that JJs with SOC and magnetism hold great promise for JDs. 

Superconductors with SOC under magnetic fields have also been explored for realizing topological superconductivity \cite{RevModPhys.83.1057,tanaka2011symmetry,leijnse2012introduction,beenakker2013search,sato2017topological,lutchyn2018majorana,prada2019andreev,frolov2019quest,flensberg2021engineered,Marra_2022}, a topological state  that hosts Majorana bound states (MBSs) and promises to revolutionize future quantum technologies. MBSs emerge above a critical magnetic field as charge-neutral topologically protected quasiparticles and exhibit spatial nonlocality \cite{PhysRevLett.100.096407,PhysRevLett.103.020401,PhysRevLett.103.107002,PhysRevLett.105.177002,PhysRevLett.105.077001}, properties that have been recently explored in JJs \cite{PhysRevLett.108.257001,PhysRevB.91.024514,PhysRevB.94.085409,PhysRevB.96.205425,cayao2018andreev,PhysRevB.104.L020501,PhysRevB.105.054504,baldo2023zero,tanaka2024theory} and shown to offer a solid way for topological qubits \cite{sarma2015majorana,Lahtinen_2017,beenakker2019search,aguado2020majorana,aguado2020perspective}.   The spatial nonlocality occurs as a result of MBSs emerging spatially separated, which allows to store information in a nonlocal manner and     immune against local sources of decoherence \cite{sarma2015majorana}.
While the detection of MBSs has recently attracted a lot of interest \cite{lutchyn2018majorana,prada2019andreev,frolov2019quest,flensberg2021engineered}, the realization of JDs in topological Ss  with MBSs has received very limited attention so far   \cite{PhysRevB.106.224509,PhysRevB.106.214524,fu2023fieldeffect,legg2023parity,lu2023tunable,cuozzo2023microwavetunable}.  In particular, it is still unknown how JDs respond to the Majorana nonlocality, an open question that could establish the   realization and use of JDs for topological quantum phenomena.

In this paper we consider S-normal-S (SNS)  JJs with SOC and Zeeman fields, and discover the emergence of highly tunable JDs, which  acquire quality factors that  are  greatly enhanced by MBSs, specially, when they become more nonlocal. We find that these topological JDs occur when the middle N region has a Zeeman field component  parallel to the SOC,  which  induces an asymmetric phase-dependent Andreev spectrum that gives rise to  supercurrents  with non-reciprocal behaviour defining the JDs. While JDs can occur even in the trivial regime, it is only in the topological phase that  they exhibit a strong dependence on the Majorana nonlocality, thus establishing the potential of MBSs for designing  JDs with topologically protected and enhanced properties.


\textit{JJs based on nanowires}.---We consider  SNS JJs formed on a single channel   nanowire with Rashba SOC [Fig.\,\ref{Fig0}(a)], with a continuum model given  by 
\begin{equation}
\label{Eq1}
H=\xi_{p_x}\tau_{z}+\frac{\alpha}{\hbar}p_{x}\sigma_{y}\tau_{z}+\Delta(x)\sigma_{y}\tau_{y}+H_{\rm Z}(x)\,,
\end{equation}
where $\xi_{p_x}=p_{x}^2/(2m)-\mu$ is the kinetic part, $p_{x}=-i\hbar\partial_{x}$ is the momentum operator, $\mu$ is the chemical potential, $\alpha$ is the Rashba SOC strength, $H_{\rm Z}(x)$ is   the space dependent Zeeman field,  $\Delta(x)$ the induced space dependent $s$-wave pair potential, and  $\sigma_{i}$ and $\tau_{i}$  are the $i$-th Pauli matrices in spin and electron-hole spaces, respectively.  For computational purposes,   Eq.\,(\ref{Eq1}) is discretized into a tight-binding lattice with spacing $a=10$\,nm  and then divided into three regions (left/right S and middle N)  of finite lengths $L_{\rm S,N}$ \cite{cayao2018andreev,baldo2023zero}, see Fig.\,\ref{Fig0}(a). The S  regions have a finite pair potential $\Delta$ with a  phase difference $\phi$, while     N   has $\Delta=0$,  originating a  SNS JJ.  To ensure that JD effect and MBSs can coexist, the Zeeman field  is taken as $H_{\rm Z}(x)=\mathbf{B}_{\rm S(N)}\cdot{\bf \Sigma}$, where ${\bf \Sigma}=(\sigma_{x}\tau_{z},\sigma_{y},\sigma_{z}\tau_{z})$, ${\bf B}_{\rm S}=(B,0,0)$, and ${\bf B}_{\rm N}=B_{\rm N}({\rm sin}\theta\,{\rm cos}\eta,{\rm sin}\theta{\rm sin}\eta,{\rm cos}\theta)$ \footnote{The Zeeman field in N can be induced by coupling N to distinct ferromagnets \cite{liu2019semiconductor,vaitiekenas2021zero,escribano2022semiconductor,PhysRevB.105.L041304,razmadze2022supercurrent}, by depositing distinct magnetic atoms \cite{nadj2014observation,PhysRevLett.121.196803,liebhaber2022quantum,trahms2023diode}, or by  a 3D vector magnet attached only to   N \cite{mourik2012signatures,wang2023triplet,PhysRevX.13.031031,PhysRevLett.125.116803}.}, with   $\theta\in(0,\pi)$ and $\eta\in(0,2\pi)$.  Moreover, we   consider realistic  parameters, with $\alpha_{\rm R}=40$\,meVnm and $\Delta=0.5$\,meV, according to experimental values reported for InSb and InAs nanowires and Nb and Al Ss \cite{lutchyn2018majorana}.

 The role of the Zeeman field can be already seen in the normal state, by inspecting the bands of Eq.\,(\ref{Eq1}) with $\Delta=0$ and  $\mathbf{B}_{\rm N}$. While a   component of  ${\mathbf B}_{\rm N}$ perpendicular to the SOC   opens a gap at zero momentum $k=0$ (yellow region), the band dispersion  becomes asymmetric  with respect to $k=0$ when ${\mathbf B}_{\rm N}$ has a term parallel     to   SOC, see  left panel of Fig.\,\ref{Fig0}(b). This asymmetry can be characterized by the difference between   the lowest bands at the SOC momenta $\pm m\alpha/\hbar^{2}$, denoted by  $\delta$, which gets a maximum at $\theta=\eta=\pi/2$  [Fig.\,\ref{Fig0}(b)]. Below we will see that this asymmetry is crucial for achieving non-reciprocal Andreev spectrum and  JDs in  JJs.

\begin{figure}[!t]
\centering
	\includegraphics[width=0.49\textwidth]{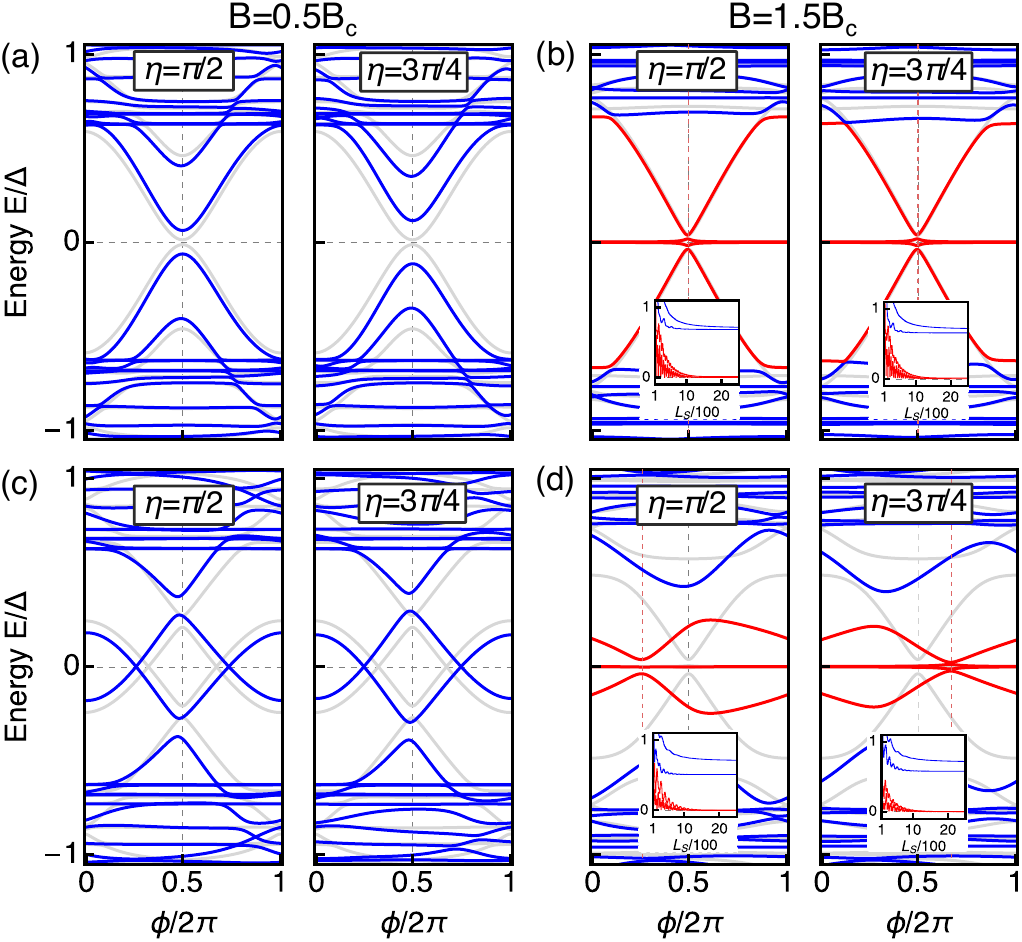}\\
 	\caption{Low-energy spectrum of JJs  with $L_{\rm N}=20$nm (a,b) and $L_{\rm N}=100$nm (c,d) as a function of $\phi$  for $B<B_{c}$ and  $B>B_{c}$  at different  $\eta$. In (b,d) the  nearest levels to zero are depicted in red color, while insets show the lowest four positive levels as function of $L_{\rm S}$. Vertical dashed lines  mark where four MBSs appear. Gray curves in all panels correspond to $\eta=0$. Parameters: $\alpha=40$meVnm, $\mu=0.5$meV, $\theta=\pi/2$, $L_{\rm S}=1000$nm, $\Delta=0.5$\,meV, $B_{\rm N}=0.5$meV.}
\label{Fig1} 
\end{figure}

\textit{Non-reciprocal phase-dependent Andreev spectrum}.---To start, we focus on the Andreev spectrum, which is presented in Fig.\,\ref{Fig1} as a function of $\phi$ for JJs with  $L_{\rm N}=20$nm  and   $L_{\rm N}=100$nm  at distinct  $\eta$ and $B$. The Andreev spectrum strongly depends on $\phi$, revealing the appearance of ABSs within the induced gap, with interesting dependences on $\eta$, $L_{\rm N}$, and $B$. At $\eta=0$, i.e., when ${\mathbf B}_{\rm N}$ is perpendicular to the SOC, the spectrum is symmetric with respect to $\phi=\pi$, see gray curves in Fig.\,\ref{Fig1}. Here,  $B=B_{\rm c}$, with $B_{\rm c}=\sqrt{\mu^{2}+\Delta^{2}}$, defines a topological phase transition into a topological phase ($B>B_{\rm c}$) with four topological ABSs that  depend on $\phi$ \cite{PhysRevLett.108.257001,PhysRevB.96.205425,PhysRevLett.123.117001}. Correspondingly, for $B<B_{\rm c}$ the system is in the trivial phase and hosts  two pairs of conventional  spin-split ABSs having a   cosine-like dependence on  $\phi$ \cite{Beenakker:92,sauls2018andreev}.  The topological ABSs  at $\phi=\pi$ define four MBSs,  two at the outer sides of S and two at their inner sides,  while the ABSs at $\phi=0$ only two MBSs  located at two outer sides of the S regions   \cite{PhysRevLett.108.257001,PhysRevB.96.205425,PhysRevLett.123.117001}. For JJs with short Ss, the four MBSs split around zero energy at $\phi=\pi$, thus giving rise to a Majorana zero-energy splitting, which gets  suppressed for long S regions and can be thus seen as a signal of the Majorana spatial nonlocality \cite{PhysRevB.104.L020501,baldo2023zero}. 

When  ${\bf B}_{\rm N}$   acquires a component that is parallel to the SOC, characterized here by $\eta\neq0$, the low-energy spectrum becomes highly asymmetric with respect to $\phi=\pi$ and develops important differences  from the $\eta=0$ case in both the topological and trivial phases [Fig.\,\ref{Fig1}(c,d)]. The asymmetry  is reflected in the ingap ABSs, which involve MBSs in the topological phase, and also in the quasicontinuum above the induced gap. For JJs with $L_{\rm N}=20$nm and $\eta\neq0$  exists only a  small asymmetry with respect to $\phi=\pi$ in the quasicontinuum but no substancial effect is seen by naked eye at low energies  [Fig.\,\ref{Fig1}(a,b)]. Note that the Majorana zero-energy splitting gets suppressed as $L_{\rm S}$ increases, consistent with their inherent spatial nonlocality, see red curves in insets of Fig.\,\ref{Fig1}(b); the ABSs for $B<B_{\rm c}$ do not depend on $L_{\rm S}$.  In JJs with $L_{\rm N}=100$nm   the Andreev spectrum exhibits a stronger response to $\eta\neq0$ [Fig.\,\ref{Fig1}(c,d)].  While the trivial spectrum here is only weakly asymmetric  [Fig.\,\ref{Fig1}(c)],   other values of $L_{\rm N}$   give spectra with larger asymmetries \cite{PhysRevB.89.195407}, see Supplemental Material (SM) \cite{SM}. Irrespective of $L_{\rm N}$, however,  the asymmetric trivial Andreev spectrum does not depend on $L_{\rm S}$ because trivial ABSs  are located only at the inner side of the JJ.  In contrast, the Andreev spectrum for $B>B_{\rm c}$ is more noticeable and strongly depends on $L_{\rm S}$  due to the presence of MBSs. In particular,  the Majorana zero-energy splitting  can occur at   $\phi$ other than $\phi=\pi$ when $\eta\neq0$  [Fig.\,\ref{Fig1}(d)], thus showing  the key role of ${\bf B}_{\rm N}$ for inducing an asymmetric Andreev spectrum in  topological JJs. Since the Majorana zero-energy  splitting  depends on $L_{\rm S}$ and   $\eta\neq0$, longer S regions give rise to  four MBSs with zero energy which then produce sharper zero-energy crossings  that are asymmetric with respect to $\phi=\pi$, see insets in Fig.\,\ref{Fig1}(d).  As a result, finite topological   JJs    host a nonreciprocal length dependent Andreev spectrum  entirely due to MBSs.

\begin{figure}[!t]
\centering
	\includegraphics[width=0.5\textwidth]{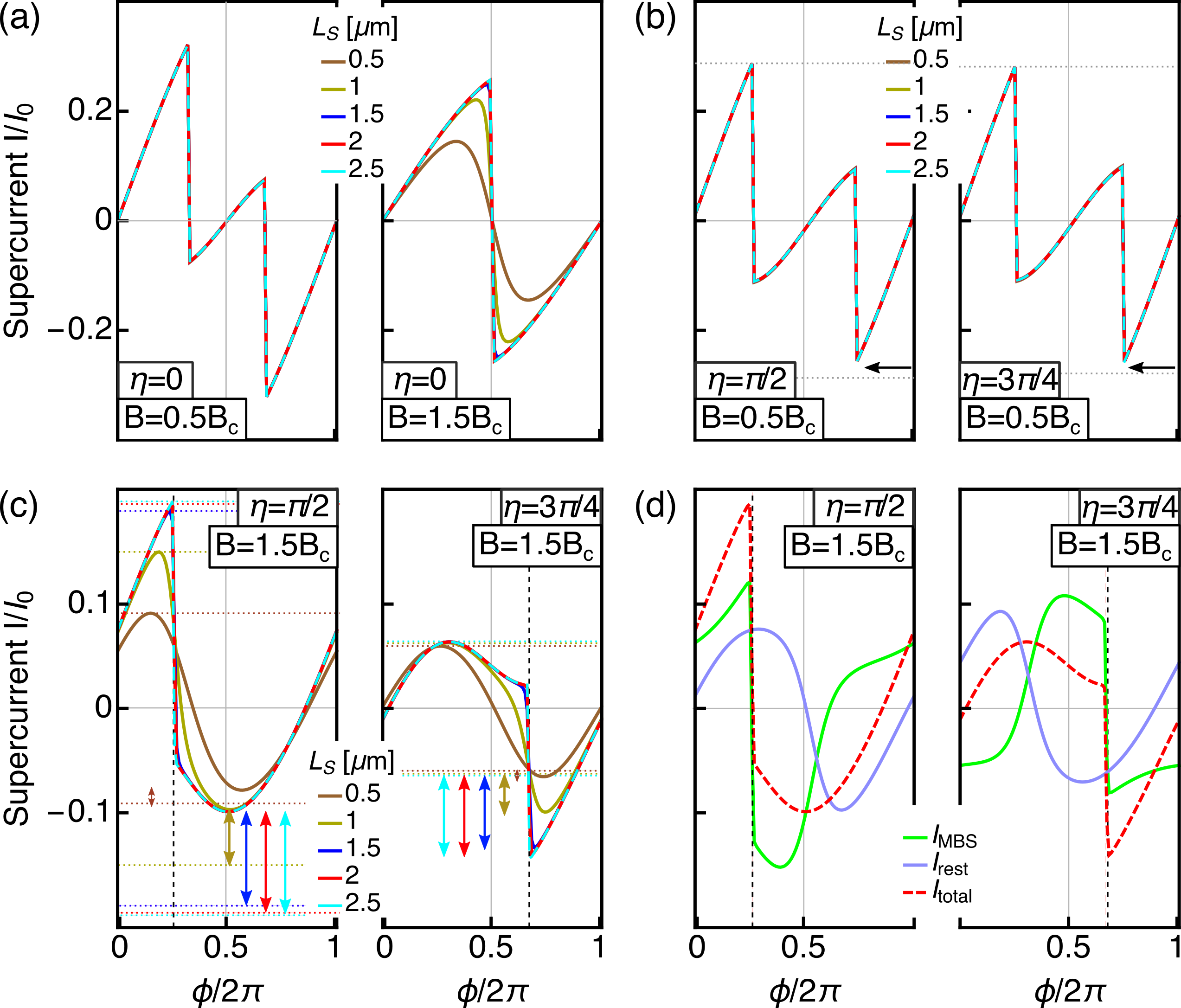}\\
 	\caption{(a) Supercurrents $I(\phi)$ in   JJs with $L_{\rm N}=100$nm as a function of   $\phi$ at
	$\eta=0$ for   $B<B_{c}$ and $B>B_{c}$ and different $L_{\rm S}$. (b,c) same as  (a) but at $\eta=\pi/2,3\pi/4$. Horizontal dotted lines mark $\pm I_{\rm c}^{+}$;   arrows in (b,d) indicate that $I^{+}_{c}\neq I^{-}_{c}$. (d) Contributions of the four MBSs and the rest of levels (ABSs and quasicontinuum) to the total $I(\phi)$ for $L_{\rm S}=2\mu$m.  Parameters: $I_{0}=e\Delta/\hbar$ and the rest as in Fig.\,\ref{Fig1}.}
\label{Fig2} 
\end{figure}

\textit{Nonreciprocal current-phase curves}.---Having established that the Andreev spectrum of  JJs  under Zeeman fields parallel to the SOC is asymmetric with respect to $\phi=\pi$, here we study how this asymmetry affects the   supercurrents $I(\phi)$. At zero temperature, we obtain $I(\phi)$ as \cite{Beenakker:92,PhysRevB.96.205425} $I(\phi)=-(e/\hbar)\sum_{\varepsilon_n>0}[d\,\varepsilon_{n}(\phi)/d\phi]$, where $\varepsilon_{n}(\phi)$ are the phase-dependent energy levels found in the previous section which include the contribution of both the ingap ABSs and the discrete quasicontinuum \cite{cayao2018andreev,PhysRevB.104.L020501}. In Fig.\,\ref{Fig2}(a-c) we present  $I(\phi)$  for   JJs with $L_{\rm N}=100$nm as a function of $\phi$ for different $\eta$ and $L_{\rm S}$  in the trivial  and topological  phases. To better understand    the role of MBSs,   panel (d) shows the contribution of MBSs ($I_{\rm MBS}$) and the rest of levels  ($I_{\rm rest}$) to  the total $I(\phi)$; here $I_{\rm rest}$ includes the contribution due to the additional ingap ABSs that coexist with MBSs and also of the quasicontinuum.

In Fig.\,\ref{Fig2}(b,c) we see that $I(\phi)$ has an overall asymmetric profile with respect to $\phi=\pi$ at $\eta\neq0$, which depends on $\eta$ and $B$, and, importantly, with a distinct response in the trivial (topological) phase to  changes in $L_{\rm S}$. This asymmetric $I(\phi)$ at $\eta\neq0$  is distinct to what is found at $\eta=0$ [Fig.\,\ref{Fig2}(a)], see also \cite{PhysRevB.96.205425,cayao2018finite,PhysRevLett.123.117001}.  The weak (strong) asymmetry in $I(\phi)$ with respect to $\phi=\pi$  stems from the  phase-dependent   Andreev spectrum  in  Fig.\,\ref{Fig1}, implying an important role of the ingap ABSs and quasicontinuum. The trivial phase with $B<B_{\rm c}$ shows a weak asymmetry for the chosen $L_{\rm N}$ due to the weakly asymmetric spectrum, but  other values can give far more asymmetric $I(\phi)$. As a result of the asymmetry, $I(\phi)$ develops a global maximum that is distinct to its global minimum, namely, different critical currents $I^{+}_{c}\neq I^{-}_{c}$, where $I^{\pm}_{c}={\rm max}_{\phi}[\pm I(\phi)]$, see black arrow in Fig.\,\ref{Fig2}(b).  Another feature of this trivial phase is that $I(\phi)$ does not change when $L_{\rm S}$ increases for any $\eta$. This insensitivity originates from that the ABSs for $B<B_{\rm c}$ emerge located at the inner sides of the JJ and  thus do not depend on $L_{\rm S}$, i.e., trivial ABSs  are not spatially nonlocal.

In contrast to the trivial phase, in the topological phase  with $B>B_{\rm c}$, $I(\phi)$ forms a larger asymmetry with respect to $\phi=\pi$  and  finite values at zero phase,  which can be  traced back to the Andreev spectrum Fig.\,\ref{Fig1}(d). Interestingly, the asymmetry of $I(\phi)$ and  $I^{+}_{c}\neq I^{-}_{c}$ strongly depends on $L_{\rm S}$, which is due to the presence of MBSs and different to what occurs for $B<B_{\rm c}$. To see $I(\phi)$ with $I^{+}_{c}\neq I^{-}_{c}$ we  indicate with colored arrows the remaining difference between the two critical currents [Fig.\,\ref{Fig2}(c)];   note that $I^{+}_{c}=I^{-}_{c}$ for $\eta=0$, as expected [Fig.\,\ref{Fig2}(a)]. Interestingly, the regimes with $I^{+}_{c}\neq I^{-}_{c}$  signal the emergence of  nonreciprocal supercurrents, or JD effect, which here occurs in the trivial  and, notably, also in the topological phases. While JDs in semiconductor-S hybrids have already been reported before  \cite{baumgartner2022supercurrent,PhysRevB.108.054522,PhysRevB.107.245415,mazur2022gatetunable,PhysRevLett.131.027001,nadeem2023superconducting,valentini2023parityconserving}, their emergence in topological JJs is intriguing because MBSs are naturally present in this regime. The effect of MBSs is evident by noting the strong dependence of  $I^{+}_{c}\neq I^{-}_{c}$  on $L_{\rm S}$ pointed out above. In fact, the topological phase  hosts four spatially nonlocal MBSs exhibiting  a zero-energy splitting at $\phi=\pi$ at  $\eta=0$ or away from $\pi$ when  $\eta\neq0$ [Fig.\,\ref{Fig1}(d)]. Then, by reducing the zero-energy Majorana splitting  with increasing $L_{\rm S}$, $I(\phi)$ acquires sharper sawtooth profiles, with larger differences between $I_{\rm c}^{+}$ and $I_{\rm c}^{-}$ that can be easily distinguished from the trivial regime. The role of MBSs can be further seen in the individual contributions of the four MBSs and the rest of levels to the total   $I(\phi)$ in  Fig.\,\ref{Fig2}(d). While  $I_{\rm rest}$ has a sizable phase dependent value, the sharpness in the  sawtooth profile of $I(\phi)$  (vertical dashed black line)  is largely determined by $I_{\rm MBS}$  due to the reduction in the Majorana zero-energy splitting for large $L_{\rm S}$. Hence, Majorana nonlocality plays a key role for enhancing the JD effect   that has not been exploited before \cite{PhysRevB.106.224509,PhysRevB.106.214524,fu2023fieldeffect,legg2023parity,lu2023tunable,cuozzo2023microwavetunable}.

\begin{figure}[!t]
\centering
	\includegraphics[width=0.5\textwidth]{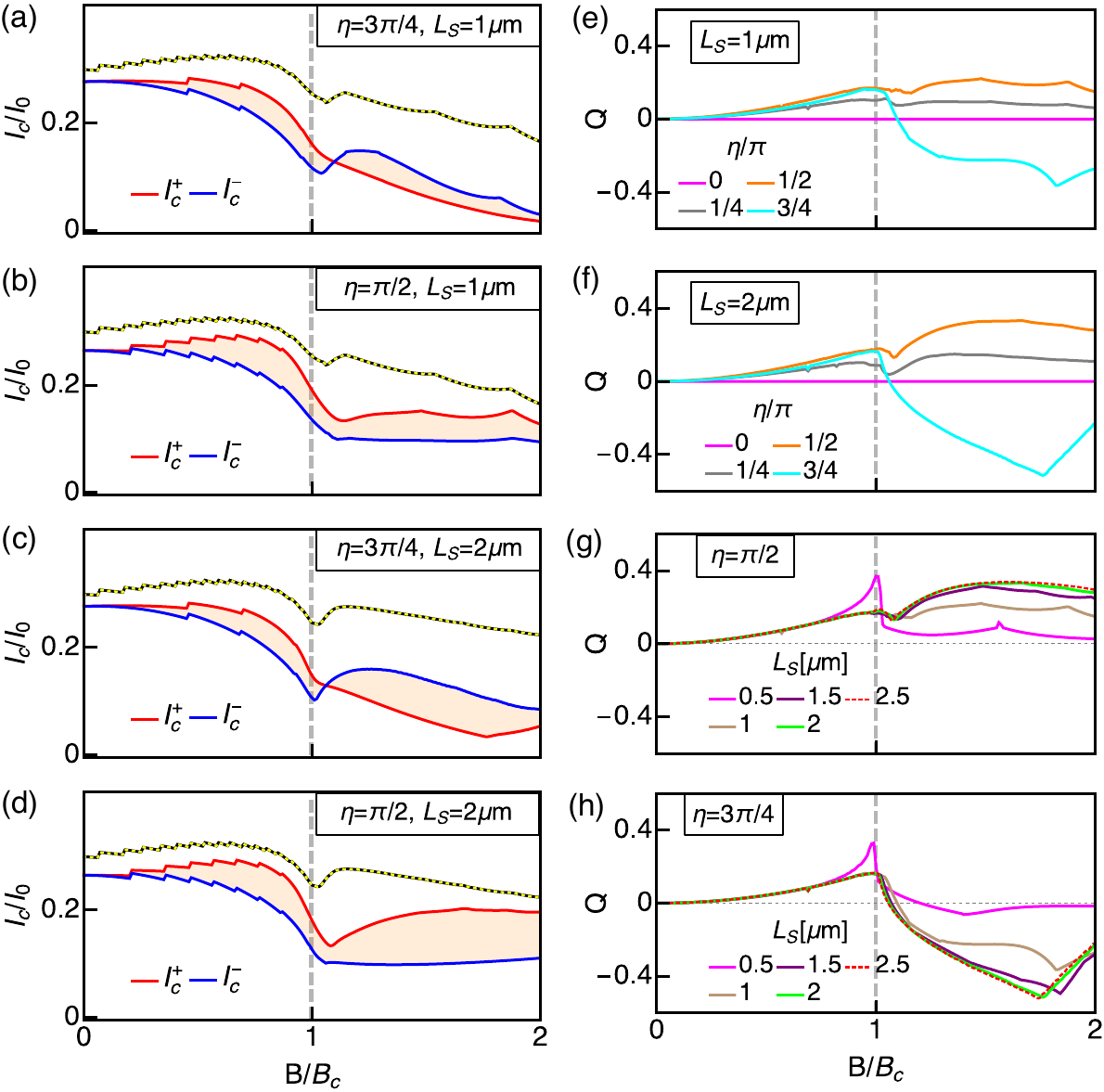}\\
 	\caption{Critical currents $I_{c}^{\pm}$ (a-d) and quality factors $Q$ (e-h) as a function of the Zeeman field  $B$ in S for different $\eta$ and  $L_{\rm S}$.  The black-yellow curve in (a-d) shows the critical current at $\eta=0$ where there is no diode effect. The vertical dashed gray line marks the topological phase transition at $B=B_{\rm c}$. Parameters:  $L_{\rm N}=100$nm and the rest as in Fig.\,\ref{Fig1}.}
\label{Fig4} 
\end{figure}

\textit{Critical currents and quality factors}.---To further understand and characterize the JDs found in previous section, here we show the critical currents $I_{\rm c}^{+}$ and $I_{\rm c}^{-}$ and their quality factors  $Q=(I_{\rm c}^{+}-I_{\rm c}^{-})/(I_{\rm c}^{+}+I_{\rm c}^{-})$. Having $Q\neq0$ reveals the amount of nonreciprocity  and a finite JD effect. In Fig.\,\ref{Fig4}(a-d) we show $I_{\rm c}^{\pm}$   as a function of $B$ for $\eta=\pi/2,3\pi/4$ and distinct $L_{\rm S}$, while in Fig.\,\ref{Fig4}(e-h) we present  $Q$ as a function of $B$. To contrast, we also plot $I_{\rm c}^{\pm}$ at $\eta=0$ in black-yellow curves of Fig.\,\ref{Fig4}(a-d). We immediately  note  that   $\eta\neq0$ induces  $I_{\rm c}^{+}\neq I_{\rm c}^{-}$, depicted by the shaded orange  regions, thus highlighting the realization of JDs. We see that the JD requires a finite $B$ in both cases $\eta=\pi/2,3\pi/4$ for the chosen $L_{\rm N}$ but in the SM we show that it can already appear at $B=0$ \cite{SM}.  

Both critical currents $I_{\rm c}^{\pm}$ reduce as  $B$ increases but $I_{\rm c}^{+}\neq I_{\rm c}^{-}$  persists   and develop a kink at $B=B_{\rm c}$, followed  by finite  values for $B>B_{\rm c}$, see  Fig.\,\ref{Fig4}(a,b). 
Increasing $L_{\rm S}$ does not change the difference between $I_{\rm c}^{\pm}$  for $B<B_{\rm c}$, but, interestingly, it does  $B>B_{\rm c}$,   inducing a larger nonreciprocity and the realization of  enhanced JDs [Fig.\,\ref{Fig4}(c,d)].  As discussed before, the sensitivity of the topological phase to changes in  $L_{\rm S}$ is because this regime hosts spatially nonlocal MBSs, whose localization and zero-energy splitting strongly depends on $L_{\rm S}$.  Hence, we can conclude that the nonreciprocity in the critical currents, and their JDs, gets enhanced entirely due to the presence of MBSs, specially, due to its spatial nonlocality.

The nonreciprocity in $I_{c}^{\pm}$    gives rise to quality factors $Q$ with a unique Zeeman dependence that confirms the impact of the  nonlocal MBSs [Fig.\,\ref{Fig4}(e-h)].  At $\eta=0$,  $Q=0$ due to   $I_{c}^{+}=I_{c}^{-}$,  as expected, see magenta lines in Fig.\,\ref{Fig4}(e,f). In contrast, at $\eta\neq0$, $Q$ gets finite values as $B$ increases, exhibiting a peak at $B=B_{\rm c}$ and finite values  for $B>B_{\rm c}$ that strongly depend on $L_{\rm S}$ [Fig.\,\ref{Fig4}(e,f)]; see also  	SM \cite{SM}. These features are absent for $B<B_{\rm c}$ and, hence,   suggest  a direct effect due to MBSs. To support this view,  in Fig.\,\ref{Fig4}(g,h) we show $Q$  as a function of $B$ at $\eta=\pi/2,3\pi/4$ and distinct $L_{\rm S}$. 
Here, $Q$ for $B<B_{\rm c}$ does not sense changes in $L_{\rm S}$ but it strongly reacts for $B>B_{\rm c}$, developing higher values  \cite{SM}.  
The peak of $Q$ at $B=B_{\rm c}$ in Fig.\,\ref{Fig4}(g,h) is due to the sharp Zeeman dependence of the ABS energies  when the gap closing signals the topological phase transition. Since for $B>B_{\rm c}$   the Majorana zero-energy splitting  strongly depends  on $L_{\rm S}$, with vanishing values for long S, the response of $Q$  to changes in $L_{\rm S}$ seen in Fig.\,\ref{Fig4}(g,h) is only attributed to the Majorana nonlocality.   Also, for certain $\eta$, $Q$   changes sign only when $B>B_{\rm c}$, showing that reversing the diode's  polarity is intriguingly related to  MBSs, see   Fig.\,\ref{Fig4}(e,f,h) \cite{SM}. Thus, topological JJs can notably realize JDs with larger quality factors due to the nonlocal nature of  MBSs.

In conclusion, we  studied  Josephson diodes in finite  topological Josephson junctions and found that their emergence is induced by having a Zeeman field in the normal region parallel and perpendicular to the spin-orbit coupling. We discovered that the quality factors of the Josephson diodes in the topological phase  are   greatly enhanced entirely due to the  nonlocality of Majorana bound states,  a mechanism that has not been explored before \cite{PhysRevB.106.224509,PhysRevB.106.214524,fu2023fieldeffect,legg2023parity,lu2023tunable,cuozzo2023microwavetunable}. Similar Josephson junctions as those studied here based on superconductor-semiconductor hybrid systems   have already been fabricated \cite{tiira17,PhysRevLett.121.047001,PhysRevX.9.011010,PhysRevLett.124.226801,PhysRevLett.125.116803,baumgartner2022supercurrent,mazur2022gatetunable,PhysRevLett.131.027001,spanton2017current}, which places our findings within experimental reach.  Our results thus establish   topological superconductivity for realizing topological Josephson diodes with protected and  enhanced functionalities.

We thank  Y. Asano, S. Ikegaya,  and S. Tamura  for insightful discussions.   
J. C. acknowledges  support from the Japan Society for the Promotion of Science via the International Research Fellow Program,  the Swedish Research Council (Vetenskapsr{\aa}det Grant No. 2021-04121),  and the Carl Trygger’s Foundation (Grant No. 22: 2093).  N. N. acknowledges support from JST CREST Grant No. JPMJCR1874, Japan.  Y. T. acknowledges support from JSPS with Grants-in- Aid for Scientific Research (KAKENHI Grants No. 20H00131 and No. 23K17668). 

\bibliography{biblio}
  \onecolumngrid



\section*{Supplementary material}

# Supplemental Material for "Enhancing the Josephson diode effect with Majorana bound states"

Jorge Cayao,[1] Naoto Nagaosa,[2] and Yukio Tanaka[3]

[1] *Department of Physics and Astronomy, Uppsala University, Box 516, S-751 20 Uppsala, Sweden*
[2] *Center for Emergent Matter Science (CEMS), RIKEN, Wako, Saitama, 351-0198, Japan*
[3] *Department of Applied Physics, Nagoya University, Nagoya 464–8603, Japan*

(Dated: January 5, 2024)

In this supplemental material we provide further supporting the our findings of the main text. In particular, below we present additional calculations of the low-energy spectrum, critical currents, and quality factors for Josephson junctions (JJ) with distinct lengths of the normal (superconducting) region $L_{\rm N(S)}$.

To begin, we remind that we here study finite superconductor-normal-superconductor (SNS) JJs with spin-orbit coupling (SOC) in the presence of Zeeman fields. The SOC is homogeneous all over the junction and oriented along $y$ axis. In contrast, the Zeeman field is inhomogeneous: in the S regions it is along $x$ and thus perpendicular to the SOC, but in the N region it has components that are parallel and perpendicular to the SOC. Thus, we denote the Zeeman field in S simply by $B$ while in N it is given by $\mathbf{B}_{\rm N} = B_{\rm N}(\sin\theta\cos\eta, \sin\theta\sin\eta, \cos\theta)$, with $\theta \in (0, \pi)$ and $\eta \in (0, 2\pi)$. Thus, by taking $\theta = \pi/2$, $\eta$ controls the Zeeman field in N to be either parallel or perpendicular to the SOC. As we discuss in the main text, having a Zeeman field parallel to the SOC creates an asymmetry in the Andreev spectrum that turns out to be essential ingredient for producing Josephson diodes in JJs. Below we provide further calculations supporting the realization of Josephson diodes in the presence of Majorana bound states (MBSs)

## CRITICAL CURRENTS AND QUALITY FACTORS FOR VERY SHORT N REGIONS

In this section we focus on JJs with middle N regions having $L_{\rm N} = 20$nm. Before going further, we point out that the Andreev spectrum of these very short JJs was presented in Fig. 1(a,b) of the main text. Here, we provide further details on the critical currents $I_c^{\pm}$ and quality factors $Q$. This is presented in Fig. S1 as a function of the Zeeman field $B$ for different values of $\eta$, which characterizes the orientation of the Zeeman field in N. For comparison, we also show the critical currents at $\eta = 0$, see dashed black-yellow curve.

As the Zeeman field $B$ increases, the critical currents $I_c^{\pm}$ first decrease and develop a kink at the topological phase transition $B = B_{\rm c}$, see Fig. S1(a,b). Above $B_{\rm c}$, the critical currents develop oscillations which stem from the Majorana zero-energy splitting seen in the Andreev spectrum. Such critical current oscillations are washed out when the S regions become longer Fig. S1(c,d), a property that reflects the spatial nonlocality of MBS. We note that both critical currents $I_c^{\pm}$ exhibit roughly the same behaviour, with very small deviations at strong $B$, close to and in the topological phase. As we discussed in the main text, these small deviations signal the occurrence of the Josephson diode effect. To characterize the diode effect, in Fig. S1(e-h) we show the quality factors as a function of the Zeeman field $B$ as $Q = (I_c^{+} - I_c^{-})/(I_c^{+} + I_c^{-})$. Here we can observe that the deviations in the critical currents $I_c^{\pm}$ produce finite quality factors, which are highly dependent on the orientation of the Zeeman field in N.

An intriguing result in this part is that the quality factors strongly depend on the length of the S regions $L_{\rm S}$ in the topological phase ($B > B_{\rm c}$) but do not in the trivial phase ($B < B_{\rm c}$). Interestingly, by increasing the length of S in the topological phase, the quality factors acquire larger values and saturate to a finite value when the four MBSs are truly zero modes, see Fig. S1(g,h). In the trivial phase, in contrast, the quality factors do not change when $L_{\rm S}$ varies. As explained in the main text, this effect stems from the fact that only the topological phase hosts nonlocal states, the MBSs, which are located at the ends of the S regions and therefore depend on the length of the S regions $L_{\rm S}$. This length dependence in the topological phase forces the zero-energy Majorana splitting seen in the spectrum in Fig. 1(a,b) to reach zero energy for long S regions, which then gives rise to larger quality factors. We also point out that there are points where the quality factor changes sign, a phenomenon that reverses the diode polarity and occurs only in the topological phase, thus suggesting an intriguing relationship between the quality factor sign change and the formation of MBSs. Furthermore, it is worth noting that the quality factors in Fig. S1(g,h) are smaller than for the JJs with $L_{\rm N} = 100$nm shown in Fig. 4 of the main text. Nevertheless, the positive impact of MBSs, that they increase the quality factors in the topological phase, remains, thus supporting our findings of the main text.

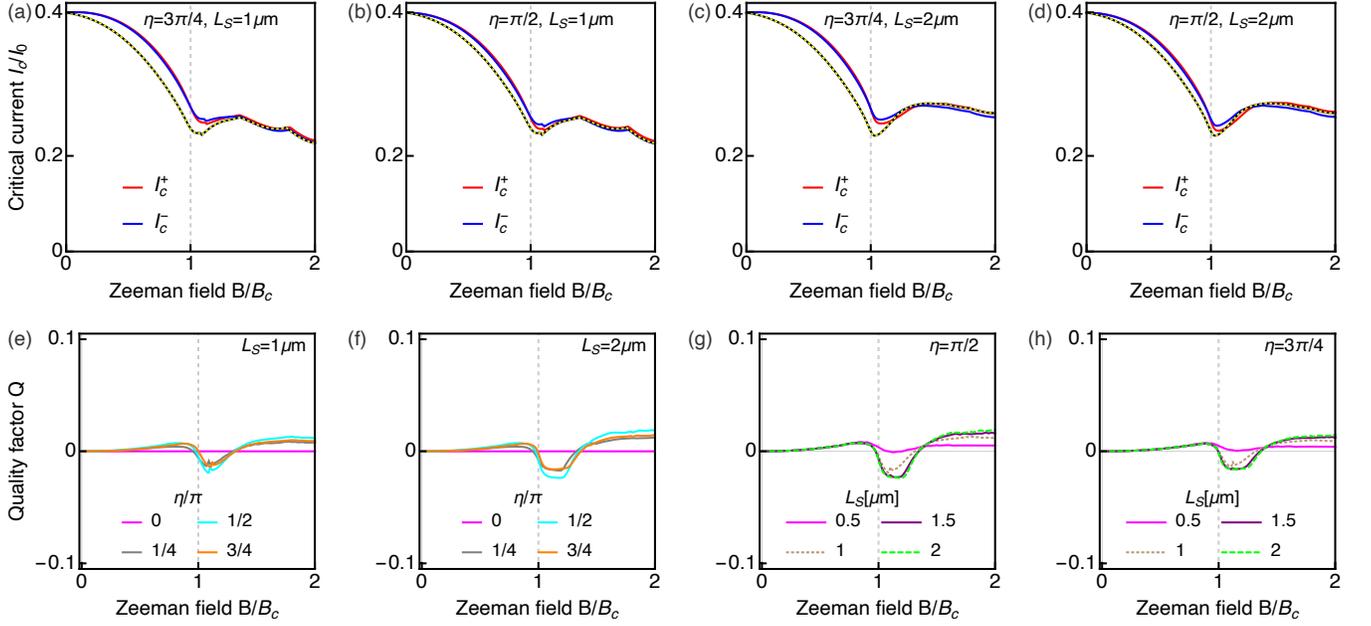

FIG. S1. Critical currents $I_c^\pm$ (a-d) and quality factors $Q$ (e-h) as a function of the Zeeman field $B$ in Josephson junctions with $L_N = 20$nm for different $\eta$ and $L_S$. The black-yellow curve in (a-d) shows the critical current at $\eta = 0$ where there is no diode effect. The vertical dashed gray line marks the topological phase transition at $B = B_c$. Parameters: $\alpha = 40$meVnm, $\mu = 0.5$meV, $\theta = \pi/2$, $\Delta = 0.5$ meV, $B_N = 0.5$meV, $I_0 = e\Delta/\hbar$.

## ANDREEV SPECTRUM, CRITICAL CURRENTS, AND QUALITY FACTORS FOR N REGIONS WITH INTERMEDIATE LENGTHS

In this section we focus on JJs with middle N regions having $L_N = 500$nm. First, in Fig. S2 we show the low-energy spectrum as a function of the phase difference $\phi$ in the trivial ($B = 0.5B_c$) and topological phases ($B = 1.5B_c$) and different $\eta$ and $L_S$. Then, in Fig. S3 we present the critical currents $I_c^\pm$ and quality factors $Q$ as a function of the Zeeman field $B$ for different $\eta$ and $L_S$.

With respect to the phase-dependent Andreev spectrum, the first observation is that it does not depend on the length of the S regions $L_S$ in the trivial phase, see top row of Fig. S2; the only effect of $L_S$ in this regime is that it adds more levels to the quasicontinuum, which is above the induced gap. The spectrum, however, depends on the orientation of the Zeeman field in N, via $\eta$ and seen, for instance, by comparing panels (a,c) with (b,d). This dependence induces an asymmetry around $\phi = \pi$ in both the ingap Andreev bound states and the quasicontinuum above the induced gap. In the topological phase, however, the situation is different, see bottom row of Fig. S2. While short S regions develop a finite Majorana zero-energy splitting, such zero-energy splitting is considerably reduced in long S regions, leading to MBSs with zero energy, see Fig. S2(e-h) and vertical dashed red lines. This dependence on the length of the S regions stems from the fact that the topological regime hosts spatially localized MBSs, as discussed in previous section and also in the main text. Furthermore, we note that, besides the effect of $L_S$, the spectrum in the topological phase also depends on the orientation of the Zeeman field in N. It becomes asymmetric with respect to $\phi = \pi$, and has regimes with the Majorana zero-energy splitting away from $\pi$. In sum, the spectrum in the trivial phase does not depend on $L_S$ but, interestingly, the Andreev spectrum in the topological phase, accompanied by the Majorana zero-energy splitting, strongly depend on the length of the S regions.

The properties of the Andreev spectrum determine the profile of the critical currents $I_c^\pm$ and quality factors $Q$, as we can see in Fig. S3. The very first feature in the critical currents is that a $\eta \neq 0$ induces $I_c^+ \neq I_c^-$, which can be clearly seen in the orange region between blue and red curves. This feature shows the emergence of the Josephson diode effect. Note that at $\eta = 3\pi/4$, the JJ exhibits a diode effect already at $B = 0$, which persists even for $B > B_c$ but with a reduced size in short S regions [Fig. S3(a)]. The critical currents $I_c^\pm$ also develop oscillations above $B_c$, which occurs due to the Majorana zero-energy splitting revealed in the phase-dependent spectrum [Fig. S2(e,f)]; At $\eta = 0$, both $I_c^\pm$ exhibit the same behaviour as a function of $B$, see black-yellow dashed curves. Moreover, for $\eta = \pi/2$ the considered JJ at $B = 0$ has $I_c^+ = I_c^-$ and thus no sign of a diode effect but at large $B$ the JJ realizes $I_c^+ \neq I_c^-$, which, interestingly, induces a finite diode effect [Fig. S3(b)]. Larger S regions gives a larger difference between $I_c^+$

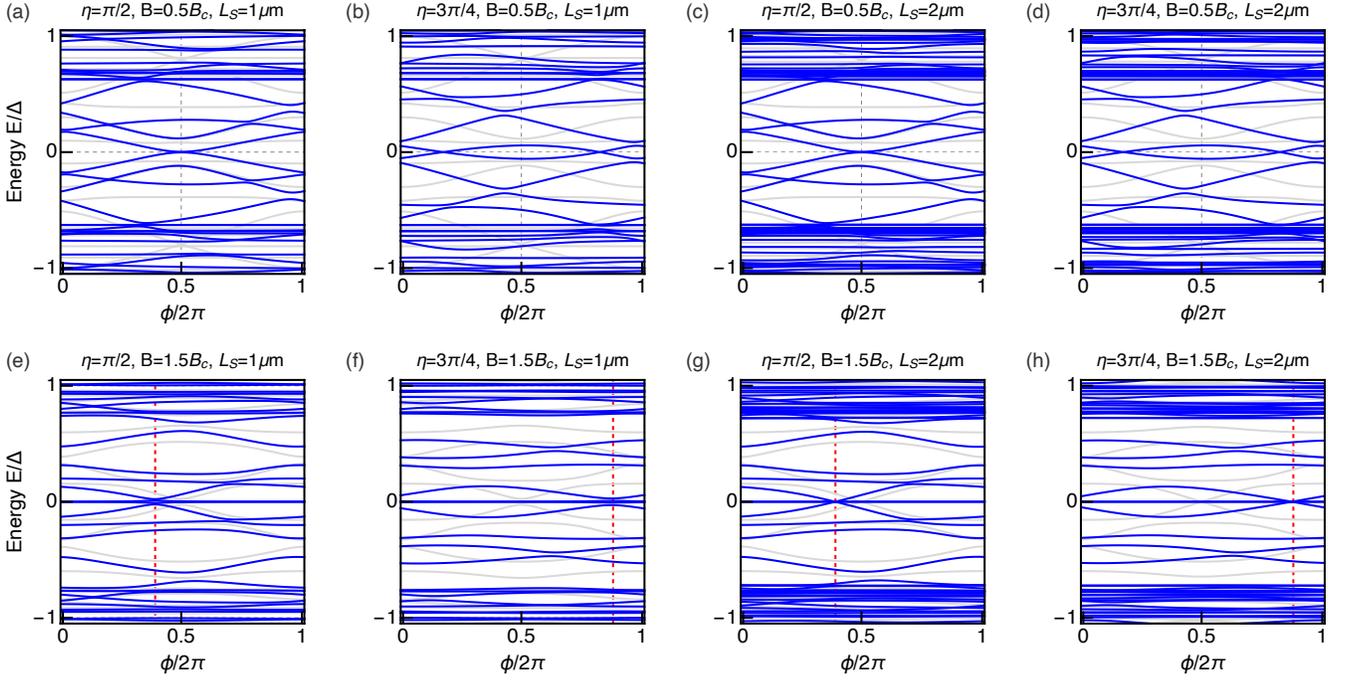

FIG. S2. Low-energy spectrum as a function of $\phi$ in Josephson junctions with $L_N = 500$nm and different $\eta$ in the trivial (a-d) and topological (e-h) phases. Gray curves in all panels correspond to $\eta = 0$. Vertical red dashed lines mark where four MBSs appear and their zero-energy splitting in the limit of large $L_S$. Parameters: $\alpha = 40$meVnm, $\mu = 0.5$meV, $\theta = \pi/2$, $\Delta = 0.5$ meV, $B_N = 0.5$meV.

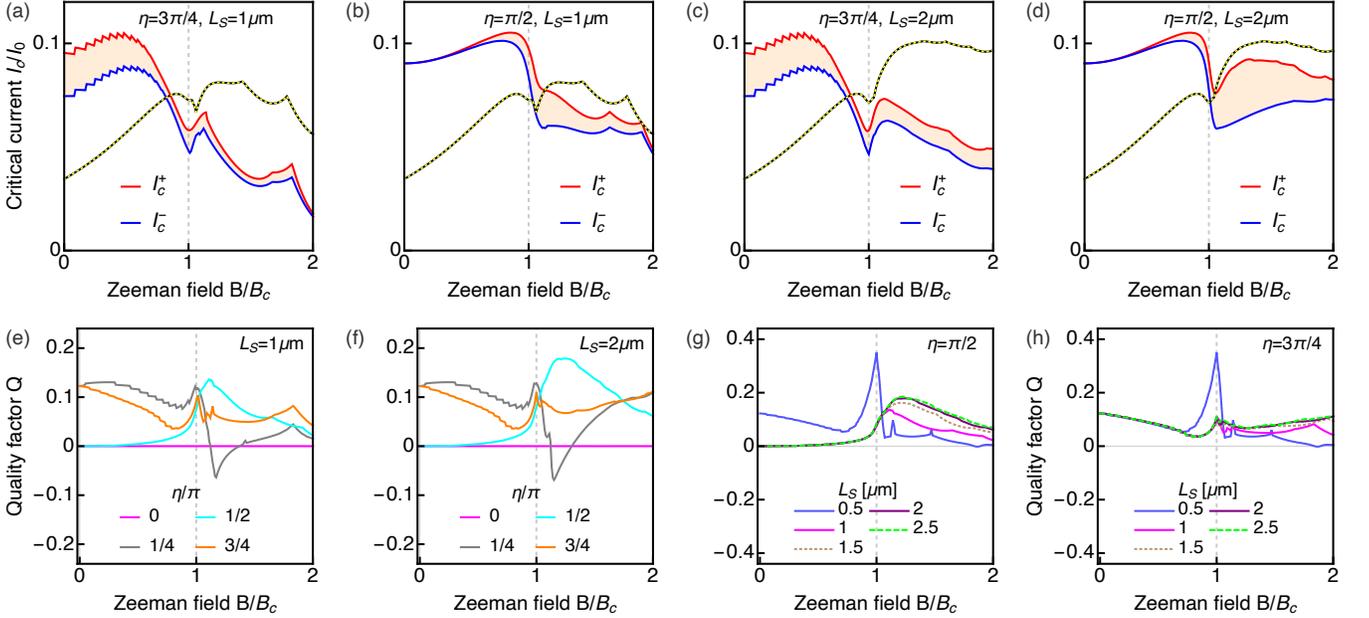

FIG. S3. Critical currents $I_c^{\pm}$ (a-d) and quality factors $Q$ (e-h) as a function of the Zeeman field $B$ in Josephson junctions with $L_N = 500$nm for different $\eta$ and $L_S$. The black-yellow curve in (a-d) shows the critical current at $\eta = 0$ where there is no diode effect. The vertical dashed gray line marks the topological phase transition at $B = B_c$. Parameters: $\alpha = 40$meVnm, $\mu = 0.5$meV, $\theta = \pi/2$, $\Delta = 0.5$ meV, $B_N = 0.5$meV, $I_0 = e\Delta/\hbar$.

and $I_c^-$, which can be seen to produce a larger Josephson diode effect [Fig. S3(c,d)]. As discussed in the main text, the dependence of the critical currents on $L_S$ in the topological phase is due to the presence of MBSs.

The Josephson diode effect can be further seen in their quality factors, presented in the bottom row of Fig. S3.

As expected due to the discussion in previous paragraph, the regime with $\eta = 3\pi/4$ exhibits a finite quality factor already at $B = 0$ diode effect, which persists at finite $B$ and even reveal the Majorana oscillations in the topological phase. At $\eta = \pi/2$ a finite quality factor appears at strong $B$ as $B$ approaches the topological phase transition at $B_c$: It develops a maximum after $B_c$ and also reflects the Majorana oscillations for $B > B_c$, see cyan curves in Fig. S3(e,f). Note that these quality factors are smaller than those found for $L_N = 100$nm, showing the dependence of the diode effect on $L_N$. By exploring the dependence on $L_S$ we find that the quality factors in the topological phase get considerably enhanced as $L_S$ increases Fig. S3(g,h), effect that can be only attributed to the presence of MBSs. This, again, supports our main finding, that MBSs enhance the Josephson diode effect, discussed in the main text. To end this part, we note that our findings can also help distinguishing between MBSs and topologically trivial zero-energy states, which is possible because the mechanism to enhance the diode effect we report here depends on MBSs.

## IMPACT OF THE LENGTH OF THE S AND N REGIONS ON THE QUALITY FACTORS

In this section we we analyze the dependence of the quality factors $Q$ on $L_S$ and $L_N$. In particular, in Fig. S4 we show the quality factors in the trivial and topological regimes as a function of $L_S$ and $L_N$ for $\eta = \pi/2, 3\pi/4$. We consider two characteristic values of the Zeeman field for the trivial ($B = 0.5 B_c$) and topological ($B = 1.5 B_c$) phases but our conclusions are valid for all $B$ in such trivial and topological regimes.

In the trivial phase ($B < B_c$), the quality factors do not depend on $L_S$, see magenta curves in Fig. S4(a-f), irrespective of the length of the N region and the value of $\eta$. As exposed in previous sections and also in the main text, this occurs because the trivial regime does not host spatially nonlocal states. In contrast, in the topological phase ($B > B_c$) the quality factors are affected by $L_S$: they increase as $L_S$ acquires larger values and saturate when the four MBSs acquire zero energy, see brown curves in in Fig. S4(a-f). This therefore makes MBSs to be important for enhancing the Josephson diode effect in finite topological JJs. Furthermore, we note that the quality factors exhibit an oscillatory behaviour as a function of $L_N$ in both the trivial and topological phases, an effect that stems from the finite size of N and is thus not of topological origin. For relatively short N regions, the quality factors in the topological phase can reach higher values than those in the trivial phase. In sum, we find that the spatial nonlocality of MBSs plays a key role for enhancing the Josephson diode effect, as argued in the main text.

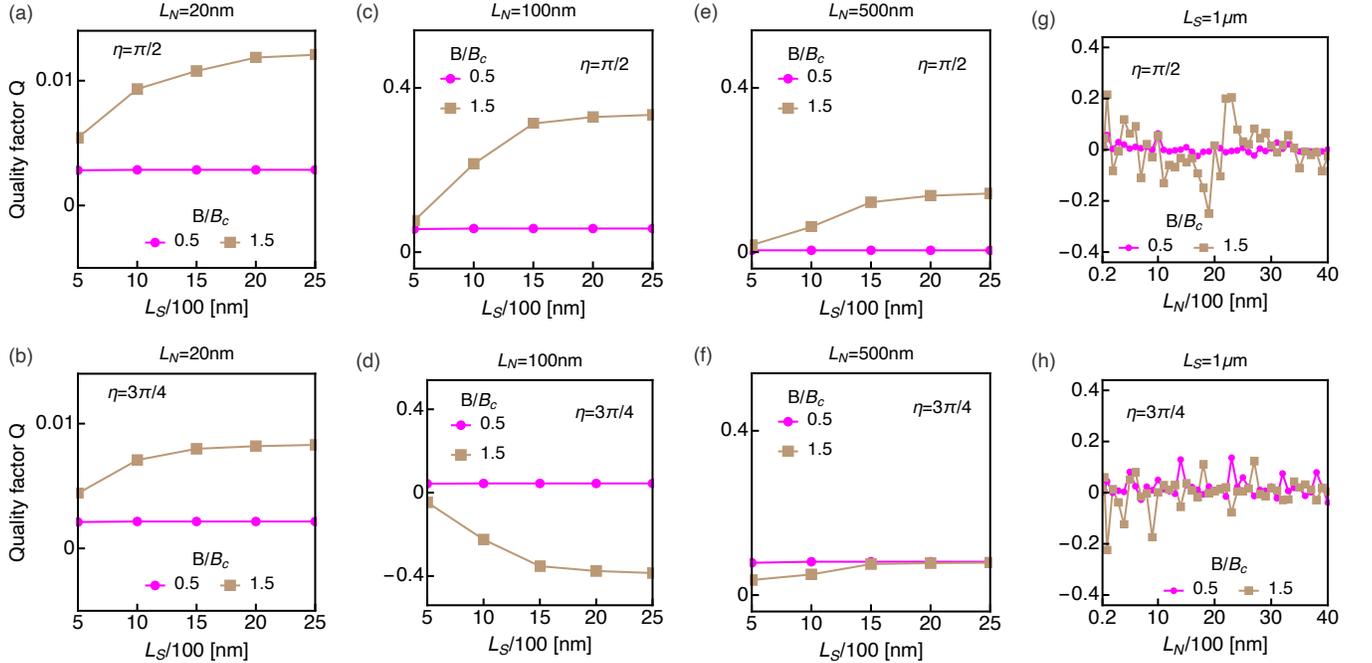

FIG. S4. Critical currents as a function of $L_S$ (a-f) and as a function of $L_N$ (g,h) in the trivial ($B < B_c$) and topological phases ($B > B_c$) for $\eta = \pi/2, 3\pi/4$. Parameters: $\alpha = 40$meVnm, $\mu = 0.5$meV, $\theta = \pi/2$, $\Delta = 0.5$ meV, $B_N = 0.5$meV, $I_0 = e\Delta/\hbar$.


\end{document}